\documentstyle[preprint,aps]{revtex}

\newcommand{\be}{\begin{equation}}
\newcommand{\ee}{\end{equation}}
\newcommand{\bea}{\begin{eqnarray}}
\newcommand{\eea}{\end{eqnarray}}
\newcommand{\nn}{\nonumber}

\newcommand{\ps}{\partial\!\!\!/}
\newcommand{\og}{\overline{G}}

\begin{document}
\draft
\preprint{YUMS 94-04, SNUTP 94-44}

\title{Gaussian approximation of the (2+1) dimensional \\Thirring model
in the functional Schr\"{o}dinger picture}

\author{
Seungjoon Hyun\footnote{e-mail: hyun@phya.yonsei.ac.kr},
Geon Hyoung Lee\footnote{e-mail: ghlee@phya.yonsei.ac.kr},
and
Jae Hyung Yee\footnote{e-mail: jhyee@phya.yonsei.ac.kr}
}
\date{April, 1994}

\address{ Department of Physics\\
     Yonsei University\\
     Seoul 120-749, Korea}

\maketitle

\begin{abstract}
The (2+1)-dimensional Thirring model is studied by using the Gaussian
approximation method
in the functional Schr\"odinger picture.
Although the dynamical symmetry breaking does not occur in the large N limit,
it does occur in the Gaussian approximation which includes the higher order
contributions in 1/N.
\end{abstract}


\newpage

\section{Introduction}
\indent
The Gaussian variational method in the functional Schr\"odinger-picture
has been shown to be
useful for the study of quantum structures of field theories \cite{gen}.
This method has been shown to be well-suited not only for the study of
bosonic field theories, but also for the study of fermionic field
theories\cite{jhy2}.
The Gaussian method is non-perturbative from the viewpoint of
both the ordinary weak coupling expansion and
also the 1/N expansion, and hence it can give the better informations
than the large-N expansion in principle, although it has not been realized yet.
It is the purpose of this paper to give an example
where the Gaussian approximation method provides better information
than the large N approximation.

In the Floreanini-Jackiw representation the fermion field operators\cite{flo},
which satisfy  the anticommutation relations
\be
\{ \psi_i(x,t),\psi_j^\dagger(y,t) \}
= i \delta_{ij} \delta(x,y),
\ee
are realized as
\bea
\psi(x)=\frac{1}{\sqrt{2}}[u(x)+\frac{\delta}{\delta u^\dagger (x)}],
 \nn \\
i \psi^\dagger (x)=\frac{i}{\sqrt{2}}[u^\dagger (x)+\frac{\delta}{\delta u(x)}]
\eea
with anticommuting Grassmann variables $u$ and $u^{\dagger}$.
For the variational approximation, we take the trial wavefunctional
in the Gaussian form
\bea
\mid \Psi > \rightarrow
\mid G >= \frac{1}{(det\,G)^{1/4}}\,
exp[\int_{x,y} u^\dagger (x)\;G(x,y)\;u(y)],
\label{gau1}  \\
< \Psi \mid \rightarrow
< G  \mid = \frac{1}{(det\,\overline{G})^{1/4}}\,
exp[\int_{x,y} u^\dagger (x)\;\overline{G}(x,y)\;u(y)],
\label{gau2}
\eea
where
\be
\overline{G}=(G^\dagger)^{-1}.
\label{con1}
\ee

This prescription has been used in \cite{jhy2} to study the Gross-Neveu model.
In the present paper, we apply this Gaussian variational method to study
some aspects of the $(2+1)$-dimensional Thirring model.
In the models considered so far, the Gaussian variational method gives
qualitatively the same results as the large $N$ approximation.
We will show that the Gaussian approximation method applied to
the (2+1)-dimensional Thirring model
provides new informations that can not be obtained
in the large $N$ limit.

The massless Thirring model has, at the classical level,
parity and chiral symmetry in two and four component representations of
fermions, respectively.
Recently it has been shown\cite{gomes,park} that in this model,
dynamical breaking of parity (chiral) symmetry occurs
as a cooperative effect of different orders in $1/N$ expansion.
In this paper we consider this problem in the context of Gaussian
approximation scheme.
It is shown that the symmetry breaking occurs in the Gaussian approximation
although it does not in the large N limit.
This agrees qualitatively with the results of \cite{gomes,park}.

\section{Dynamical symmetry breaking in two-component representation}
\subsection{Schr\"odinger picture Gaussian approximation}
The (2+1)-dimensional  massless Thirring model is described by the Lagrangian
density
\be
{\cal L}_0 = i \overline{\psi}_a \ps \psi_a
         -\frac{g}{2N}(\overline{\psi}_a \gamma^\mu\psi_a)
         (\overline{\psi}_b \gamma_\mu\psi_b),
\label{lag}
\ee
where $a=1,2,\cdot \cdot \cdot,N.$
In (2+1) dimensions the $\gamma$ matrices for two-component representations
of fermions can be represented by the $2 \times 2$ Pauli matrices,

\be
\gamma^0=\sigma^3,\,\,\,\,\,
\gamma^1=i \sigma^1,\,\,\,\,\,
\gamma^2=i \sigma^2.
\ee
As is well-known, the model has parity invariance at the classical level,
under which the fermions transform as
\be
\psi(x^0,x^1,x^2) \rightarrow \gamma^1 \psi(x^0,-x^1,x^2).
\ee
The mass term $\overline{\psi} \psi$ changes sign under the transformation and
thus breaks the symmetry.

In order to compute the effective potential and see if dynamical symmetry
breaking occurs, we introduce the vacuum condensate of fermion bilinear,
\be
\sigma_a \equiv -\frac{g}{2}<\overline{\psi}_a \psi_a>,
\ee
where $a$ is a color index,
and compute the effective potential as a function of $\sigma_a$.
One can achieve this by writing the Lagrangian as
\be
{\cal L} \equiv {\cal L}_0
+\alpha_a(\sigma_a+\frac{g}{2}\overline{\psi}_a \psi_a).
\ee

By using the Gaussian trial wave functional (\ref{gau1}) and (\ref{gau2}),
we calculate the vacuum expectation value of the Hamiltonian of the model,
\bea
<\!G\!\! \mid \!\!H \!\!\mid\!\!G\!>&=&\frac{1}{2} \int \!\!d^2\!x d^2\!y\,
Tr[(-i \gamma^0 \gamma^i \partial_i(x,y)
+\frac{g}{2N} \delta(x,y)\gamma^0 \gamma^\mu \gamma^0 \gamma_\mu
+\frac{g}{2} \alpha_a(x) \delta(x,y) \gamma^0 )\Omega(y,x)] \nn \\
&-&\frac{g}{8N}Tr[\gamma^0 \gamma^\mu
\Omega(x,y)\gamma^0 \gamma_\mu \Omega(y,x) \delta(x,y)] \nn \\
&+& \int d^2\!x \{ \frac{g}{8N} Tr[\gamma^0 \gamma^\mu \Omega(x,x)]
Tr[\gamma^0 \gamma_\mu \Omega(x,x)] +\alpha_a(x) \sigma_a(x) \},
\label{ghg}
\eea
where
\bea
&\Omega&(x,y) \equiv 2 <\!G \!\!\mid \!\psi \psi^\dagger \!\mid\!\! G\!>=
<\!x\!\! \mid\! (1+G) S^{-1}(1+\overline{G})\! \mid\!\! y\!>, \\
&S& \equiv G + \overline{G},
\label{defomega} \\
&\partial&_i(x,y) \equiv \frac{\partial}{\partial x^i} (\delta(x^i-y^i)),
\eea
and $Tr$ denotes the trace over Dirac spinor and color indices.
In order to calculate the Gaussian approximation,
it is convenient to introduce the current $A_\mu \equiv \frac{g}{2\sqrt{N}}
<\overline{\psi} \gamma_\mu \psi>$, which can be realized by introducing the
Lagrange's multiplier into Hamiltonian,
\bea
H_{eff} &\equiv& \frac{1}{2} \int d^2x d^2y Tr[h_\Omega(x,y) \Omega(y,x)] \nn
\\
&+&\int d^2x \{ \frac{1}{2g} A^\mu(x)A_\mu(x)
+\beta_\mu(A^\mu(x)-\frac{g}{2\sqrt{N}}Tr[\gamma^0 \gamma^\mu \Omega(x,x)])
+ \alpha_a \sigma_a  \}
\label{heff}
\eea
where
\bea
h_\Omega(x,y) &\equiv& -i \gamma^0 \gamma^i \partial_i(x,y) \nn \\
&+&\frac{g}{4N} \gamma^0 \gamma^\mu(2I(x,y)-\Omega(x,y)) \delta(x,y) \gamma^0
\gamma_\mu
+\frac{g}{2} \alpha_a(x) \delta(x,y) \gamma^0.
\label{ho}
\eea
The effective potential is obtained by minimizing $H_{eff}$ with respect to
the kernel $G(x,y)$.

\subsection{Effective Potential in the Large N Limit}
Since the second term in the $h_\Omega$ of Eq(\ref{ho}) is of $1/N$ order,
the effective potential in the large N limit is obtained by considering
\be
H_{eff}=\frac{1}{2} Tr h_N \Omega+\frac{1}{2g} A^\mu A_\mu+\beta^\mu A_\mu
+\alpha_a \sigma_a,
\ee
where
\be
h_N \equiv -i \gamma^0 \gamma^i \partial_i(x,y)
+\frac{g}{2} \alpha_a(x) \delta(x,y) \gamma^0
-\frac{g}{2\sqrt{N}} \beta_\mu(x) \gamma^0 \gamma^\mu \delta(x,y) .
\ee
In order to obtain the effective potential of the system,
we take variations on $<G \mid\! \!H\!\! \mid \overline{G}>$ with respect
to $G$ and $\overline{G}$ .
The invariance of $<G \mid\!\! H \!\!\mid \overline{G}>$ under these
variations yields the equations
\bea
(I-G) h_N (I+G)=0
\label{var1}\\
(I+\og) h_N (I-\og)=0.
\label{var2}
\eea
These two conditions  are shown to be equivalent
under the condition of (\ref{con1}),
and can be solved by the well-known method \cite{jhy2}.
The Eq.(\ref{var1}) can be rewritten in the form
\be
h_N^2-K_N^2+[h,K_N]=0
\label{con2}
\ee
where $h_N$ and $K_N \equiv h_N G_N$ are $2 \times2$ matrices in the
Dirac spinor space.
Any $2 \times 2$ matrix can be expressed as a linear combination of $\Gamma^a$:
\be
\Gamma^0=I, \,\,\,\;
\Gamma^1=-i \gamma^1, \,\, \,\;
\Gamma^2=-i \gamma^2, \,\,\,\;
\Gamma^3=\gamma^0.
\ee
$K_N$ and $h_N$ is then decomposed as
\bea
K_N(x,y)&=&\sum_{a=0}^3 \Gamma^a \int \frac{d^2p}{(2\pi)^2} e^{-ip\cdot(x-y)}
K_{Na}(p)
\label{kn} \\
h_N(x,y)&=&\int \frac{d^2p}{(2\pi)^2}e^{-ip\cdot(x-y)}[-g \beta_0 \Gamma^0 \nn
\\
&-&(p_2+g \beta_2) \Gamma^1+
(p_1+g \beta_1) \Gamma^2+g \alpha_a \Gamma^3].
\label{hn}
\eea
The Eqs.(\ref{kn}),(\ref{hn}) and (\ref{con2}) yield
\bea
&K_{N0}^2&+K_{Ni} K_{Ni}=(p_1+g\beta_1)^2+(p_2+g \beta_2)^2+g^2 \alpha_a^2
+g^2 \beta_0^2, \nn  \\
&K_{N0}& K_{N1}= g \beta_0(p_2+g \beta_2)+i(p_1+g \beta_1)K_{N3}
-ig \alpha_a K_{N2},  \nn \\
&K_{N0}& K_{N2}=-g \beta_0(p_1+g \beta_1)+i(p_2+g \beta_2)K_{N3}
+ig \alpha_a K_{N1},
\label{keq} \\
&K_{N0}& K_{N3}=-g^2 \beta_0 \alpha_a -i(p_2+g \beta_2)K_{N2}
-i(p_1+g \beta_1)K_{N1},  \nn
\eea
where the summation convention for the index i(=1,2,3) is implied.
The Eq.(\ref{keq}) has the non-trivial solutions
\bea
K_{N0}&=\pm \sqrt{h_i h_i}, \nn \\
K_{Ni}&=h_0 h_i/K_{N0}
\label{ksol}
\eea
as well as the trivial solution $K_{N0}=\pm g \beta_0 $, and this gives the
solution for the kernel
\bea
G_N(x,y)&=&(h_N^{-1}K_N)(x,y)  \nn \\
&=& \pm \int \frac{d^2p}{(2\pi)^2} e^{-ip(x-y)} \frac{1}{\sqrt{h_i h_i}} h_j
\Gamma^j.
\label{gsol}
\eea

Following the similar procedure as in the case of Gross-Neveu model\cite{jhy2},
the effective potential $V_{eff}$, which is defined by
\be
H_{eff}\equiv \int d^2\!x V_{eff},
\ee
can be determined as
\bea
V_{eff}&=&N \int d^2x[
-\int \frac{d^2p}{(2\pi)^2}\{g \beta_0
+\sqrt{-(p_1+g \beta_1)^2-(p_2+g \beta_2)^2+g^2 \alpha^2} \} \nn \\
&&+\frac{A^2}{2Ng}+\frac{1}{N}\beta^\mu A_\mu+\alpha \sigma]  \\
&=&-\frac{g N}{2\pi} I_1 \beta_0-\frac{N}{2\pi}I_2
-\frac{N}{4\pi}g^2 \alpha^2 I_0(M)+\frac{N}{6\pi}g^3 \mid \alpha \mid^3 \nn \\
&&+\frac{1}{2g}A^\mu A_\mu+\beta^\mu A_\mu+N \alpha \sigma,
\label{vneff}
\eea
where
\bea
&I&_0(\!M\!) \equiv M+\int_0^\infty dp \frac{p}{\sqrt{p^2+M^2}} \nn \\
&I&_1 \equiv \int p\;dp \nn \\
&I&_2 \equiv \int p^2 \;dp.
\eea
Here we have taken the solution with minus sign
in Eq.(\ref{ksol}) and (\ref{gsol}), which corresponds to lowest energy.
Since we want to see whether dynamical symmetry breaking occurs,
the auxiliary field variables $A_\mu$ in Eq.(\ref{vneff}) should be eliminated
via
variational method.
If we neglect the irrelevant infinity, the effective potential becomes
\be
V_{eff}=-\frac{N}{4\pi} g^2 \alpha^2 I_0(M)+
\frac{N}{6\pi}g^3 \mid \!\alpha\! \mid^3 + N \alpha \sigma.
\ee
By eliminating $\sigma$, the effective potential can be written as
\be
V_{eff}=-\frac{N}{4\pi} g^2 \alpha^2 (I_0(M)-
\frac{4}{3}g  \mid \!\alpha\! \mid).
\ee
This effective potential can easily be renormalized in the form,
\be
V_{eff}=-\frac{N}{4\pi} g_r^2 \alpha^2 ,
\ee
where $g_r^2 \equiv g^2 I_0(M)$ is the renormalized coupling constant.
This implies that the symmetry breaking does not occur in the large $N$ limit.

\subsection{Beyond large N limit}
To compute the full Gaussian effective potential,
it is convenient to rewrite the original effective hamiltonian
density (\ref{heff}) in the form
\be
H_{eff}=\frac{1}{2} Tr [h\Omega]+
\frac{g}{8N} Tr [\gamma^0 \gamma^\mu \Omega \gamma^0 \gamma_\mu \Omega]+
\frac{1}{2g}A^\mu A_\mu+\beta_\mu A^\mu +\alpha_a \sigma_a
\ee
where
\be
h \equiv h_N+\frac{g}{2N} \gamma^0 \gamma^\mu(I-\Omega(x,x))
 \gamma^0 \gamma_\mu.
\label{hdef}
\ee
Taking variations on $H_{eff} $ with respect to $G$ and $\og$ yield
\be
(I-G)h(I+G)=0.
\ee
This equation has the same form as Eq.(\ref{con2}) with $h_N$ replaced by $h$.
Therefore we can solve
\be
h^2-K^2+[h,K]=0
\ee
where $K \equiv h G$.
We can follow the same procedure as in the case of large N-limit and
obtain the nontrivial solution for $K$
\be
\tilde{K}(p)=- \frac{1}{\sqrt{h_i h_i}}( h_j h_j \Gamma^0 + h_0 h_j \Gamma^j),
\ee
and
\be
\tilde{G}(p)=-\frac{1}{\sqrt{h_i h_i}}h_j \Gamma^j,
\label{gcal}
\ee
which corresponds to the minimum energy solution.
We also obtain the solution for $\Omega$, $\Omega=I+G$,
from the relations among $G,G^\dagger,K,K^\dagger,h$, and $h^\dagger$.
Then using the Eq.(\ref{hdef}) one finds
\bea
h(x,y)&=&\int \frac{d^2 p}{(2\pi)^2}
e^{-ip(x-y)}[
-\frac{g \beta_0}{2\sqrt{N}} \Gamma^0
+(-p_2+\frac{g \beta_2}{2 \sqrt{N}}-\frac{g}{2N} G_1(0))\Gamma^1 \nn \\
&+&(p_1-\frac{g \beta_1}{2\sqrt{N}}-\frac{g}{2N}G_2(0))\Gamma^2
+(\frac{g}{2}\alpha-\frac{3g}{2N}G_3(0))\Gamma^3],
\label{hcal}
\eea
where
\be
G_i(0) \equiv \int \frac{d^2p}{(2\pi)^2} \tilde{G}_i(p).
\ee
{}From Eq.(\ref{gcal}) and (\ref{hcal}), we obtain the consistency condition
\be
G_i(0)=-\int \frac{d^2p}{(2 \pi)^2} \frac{1}{\sqrt{h_j h_j}} h_i,
\ee
and therefore,
\bea
G_1(0) &=& G_2(0)=0, \nn \\
G_3(0) &=& - m \int \frac{d^2 p}{(2\pi)^2}
[(p_1-\frac{g \beta_1}{2 \sqrt{N}})^2
  +(p_2-\frac{g \beta_2}{2 \sqrt{N}})^2+m^2]^{-\frac{1}{2}} \nn \\
&=&-\frac{m}{2 \pi} (I_0-m),
\eea
where
\be
m \equiv \frac{g}{2}\alpha-\frac{3g}{2N} G_3(0).
\label{alp}
\ee
Extremizing the effective potential with respect to $G_3(0)$,
we obtain
\bea
V_{eff}&=&-\frac{N}{4\pi} I_0(M) m^2+\frac{N}{6\pi} \mid \!m \!\mid^3
+\frac{g}{12}N^2 \alpha^2 +\frac{N^2}{3g} m^2 \nn \\
&-&\frac{N^2}{3} \alpha m+N \alpha \sigma
\label{veff1}
\eea
Eq.(\ref{alp}) can then be written as
\be
\alpha-\frac{2m}{g}+{3m}{2 \pi N}(I_0-m)=0,
\ee
which can be understood as $\frac{\partial}{\partial m}V_{eff}=0$.

By eliminating $\alpha$ by  $\frac{\partial}{\partial \alpha}V_{eff}=0$
and $\sigma$ by  $\frac{\partial}{\partial \sigma}V_{eff}=0$,
the effective potential becomes
\be
V_{eff}=\frac{N}{3g}(N-\frac{3I_0}{4\pi}g)m^2+
\frac{N}{6\pi} \mid \!m\! \mid^3,
\ee
which needs to be renormalized.
$V_{eff}$ can be made finite by defining
the renormalized coupling constant $g_r$ by
\be
-\frac{1}{g_r} \equiv \frac{1}{g}(N- \frac{3I_0}{4\pi}g).
\ee
Then, the renormalized effective potential cab be written as
\be
V_{eff}=-\frac{N}{3g_r} m^2 +\frac{N}{6\pi} \mid \!m\! \mid^3.
\label{final}
\ee
This implies that the symmetry breaking occurs if $g_r \geq 0$.
Since $\sigma=N m/3$ in the extrimum, the second term of Eq.(\ref{final})
is smaller than the first term by $1/N$.
Eq.(\ref{final}) shows, in a straightforward manner,
why symmetry breaking phenomenon cannot be seen in the large N limit.

\section{Dynamical symmetry breaking in four-component representation}
To restore parity symmetry even with the explicit mass term,
we consider four-component representation of Dirac spinor.
This can be achieved by considering a doublet of two component spinors,
\bea
\psi = \left( \begin{array}{c} \psi_L \\ \psi_R \end{array} \right),
\eea
for which we define the $4 \times 4\;\; \gamma$-matrices
\bea
\gamma^1=i \Gamma_{13}=i \left( \begin{array}{cc} \sigma_1& 0 \\
0&-\sigma_1 \end{array} \right), \,\,\,
\gamma^2=i \Gamma_{23}, \,\,\,
\gamma^0=i \Gamma_{33}, \,\,\,
\gamma^5\gamma_5=\Gamma_{02}.
\eea
where $\Gamma_{ij} \equiv \sigma_i \otimes \sigma_j$.
In this representation, mass term does not spoil parity symmetry,
but break the continuous $U(1) \times U(1)$ chiral symmetry \cite{park2},
\bea
\psi_L \rightarrow e^{i \alpha_L} \psi_L, \nn \\
\psi_R \rightarrow  e^{i \alpha_R} \psi_R,
\eea
down to the $U(1)$ subgroup.

Thus the existence of mass term signals the breaking of such chiral symmetry.
We adopt the same procedure, as in two-component case, to obtain the effective
potential, and introduce two vacuum condensates of fermion bilinear
\be
\sigma \equiv -\frac{g}{2} <\overline{\psi} \psi>, \,\,\,
\pi \equiv \frac{i g}{2} <\overline{\psi} \gamma_5 \psi>.
\ee
Then the effective Hamiltonian can be written as
\be
H_{eff} = \frac{1}{2} Tr h\Omega + \frac{1}{2Ng} A^\mu A_\mu+
\beta^\mu A_\mu + \alpha_a \sigma_a+\tau_a \pi_a
\ee
where
\bea
h=-i \gamma^0 \gamma^i \partial_i(x,y)
+\frac{g}{2} \alpha(x) \delta(x,y) \gamma^0
-\frac{g}{2 \sqrt{N}} \beta_\mu \gamma^0 \gamma^\mu \nn \\
+\frac{g}{2N} \gamma^0 \gamma^\mu (1-\Omega(x,x)) \gamma^0 \gamma_\mu
-\frac{ig}{2} \tau \gamma^0 \gamma_5.
\eea
By following the same procedure as in the two component case,
we obtain the effective potential
\be
V_{eff}=-\frac{3}{2 N g_r} \phi^2+\frac{9}{8 \pi N^2} \mid \phi \mid^3
\ee
where $ \phi^2=\sigma^2+\pi^2 $.
The model exhibits the dynamical breaking of the chiral symmetry
in the exactly same manner as in the two-component case.

\section{Conclusion}
The Gaussian approximation method has been shown to be
useful in studying the non-perturbative
aspects of quantum field theories.
In the models considered so far \cite{jhy2}, the Gaussian approximation
method gives qualitatively the same results as those of large N approximation.
This is partly due to the fact that the main non-perturbative feature
of those models appears at the leading order in $1/N$ expansion.
On the other hand (2+1)-dimensional Thirring model has the dynamical
symmetry breaking not in the leading order of N,but in the combination
of different orders in $1/N$\cite{gomes,park}. In two component formalism,
this non-perturbative phenomena, from the point of view of large N
expansion, occur in the range of $g_R$,
$0 \leq g_R \leq g_c \equiv \frac{1}{16} exp(-\frac{N \pi^2}{16}) $
\cite{park}.
In this paper we use the Gaussian approximation scheme to reveal this.
We have shown that in this scheme the dynamical symmetry breaking does
occur as a cooperative effect of the leading and
next-to-leading orders in $1/N$.
Since the Gaussian approximation gives better results than the large $N$
approximation, we believe that the Gaussian approximation is well-suited
for the difficult non-perturbative problems such as the existence of
bound state.

\vspace{4cm}
\begin{center}
{\Large Acknowledgement}
\end{center}

This work was supported in part by the Ministry of Education,
Korea Research Foundation, Center for Theoretical Physics(S.N.U.) and
the Korea Science and Engineering Foundation


\end{document}